\documentclass[
 reprint,
superscriptaddress,
 amsmath,amssymb,
 aps,
pra,
tikz,border=3mm
]{revtex4-2}
\usepackage[utf8]{inputenc}
\usepackage[T1]{fontenc}
\usepackage{xfrac}
\usepackage{amsmath}
\usepackage{graphicx}
\usepackage{amsfonts}
\usepackage{hyperref}
\usepackage{amssymb}
\usepackage{physics}
\usepackage{wrapfig}
\usepackage{xcolor}
\usepackage{float}
\usepackage{indentfirst}

\begin{document}

\title{Analogue simulation of quantum gravity black hole models in a dc-SQUID array}
\author{Marco D. Maceda}
\affiliation{Departamento de F\'isica Te\'orica, Universidad Aut\'onoma de Madrid E-28049 Madrid, Spain. (marco.diaz@estudiante.uam.es)}
\author{Carlos Sab\'in}
\affiliation{Departamento de F\'isica Te\'orica and CIAFF, Universidad Aut\'onoma de Madrid E-28049 Madrid, Spain}




\begin{abstract}
\begin{center}
\textbf{Abstract}    
\end{center}

We propose an analog quantum simulation for studying the collapse and bounce of a star from infinity. In this spacetime, which encompasses both a black hole and a white hole, we place a massless scalar field that propagates at the speed of light, which is modified by the curvature. We simulate this system using an SQUID array, in which we can alter the propagation of light using an external magnetic field. We consider both infalling and outfalling radiation, giving rise to two different scenarios: downstream and upstream radiation. We compute the magnetic flux profile required by the simulation in both cases and find out that the former is more experimentally suitable.
\end{abstract}

\maketitle
%
%

\section{Introduction}
Black holes are fascinating objects from both the fundamental and the phenomenological  point of view. At classical general relativity level, they can be seen as the final result of stellar collapse, which gives rise to an event horizon and a singularity.  A semiclassical description within the framework of Quantum Field Theory in curved spacetime (QFTCS) \cite{birrell_davies_1982} already modifies this static view, giving rise to black-hole thermodynamics and the celebrated prediction of Hawking radiation \cite{hawking_radiation_Hawking1975}. However, this evaporation process leads to the question of the final fate of the black hole, which can be problematic in terms of the information loss problem \cite{PhysRevD.14.2460}. Ultimately this question cannot be fully answered in the absence of a full quantum theory of gravity, expected to govern the physics of a black hole shrank to Planck-scale sizes. For instance, in loop quantum gravity models, a resolution to the classical singularity and the information loss appears in the form of a bounce which transforms the black hole into a white hole after reaching a critical size \cite{CRStephens_1994,Barceló_2015,mutinity,PhysRevD.106.024014}. 

Due to the lack of the aforementioned full theory of quantum gravity and the practical impossibility of directly manipulating black holes, the use of Analogue Gravity \cite{barcelo2011analogue} systems appears as a resource of interest, both from the understanding of the experimental systems themselves, which are pushed to new regimes, and from the gravitational side, where the issues that arise in the experiments have an impact back in the simulated systems as well -for instance, issues on the robustness or the quantum nature of the Hawking radiation, as pointed out in \cite{barceloanalogue}. Moreover, this research path has benefited in the last years from the parallel development of Quantum Simulators \cite{daley2022practical}, namely quantum-technological setups aimed to mimic the behavior of inaccessible physical systems. Indeed, it is in the intersection of the quantum simulation and analogue gravity avenues where the celebrated observations of laboratory analogues of black holes phenomena, such as effective horizons and analogue Hawking radiation, providing valuable experimental insight complementary to astrophysical observations of real black holes\cite{boseeinstein2exp,MunozdeNova:2018fxv,Kolobov:2019qfs,Steinhauer:2021xxj}. Other modern quantum setups such as superconducting circuits have been proposed as possible platforms for analogue black holes \cite{squid6,squid4, PhysRevLett.126.041105, bekenstein2}. However, alternatives to the model of single black hole collapse are much less studied within the analogue gravity or quantum simulation communities. An exception is the work \cite{garcia2023hawking} where the authors analyze the Hawking radiation generated in an analogue black hole including a bounce, by means of an SQUID terminated coplanar waveguide. They use the equivalency between black holes and accelerated boundary conditions and propose an implementation of the latter in a superconducting circuit setup similar to the one employed in the experimental observation of the Dynamical Casimir effect, where particles were generated out of the quantum vacuum precisely by the modification of boundary conditions at large speeds \cite{wilson2011observation}. Thus, this scheme highlights the relation between Hawking radiation and the Dynamical Casimir effect. In the same spirit, there are works using the aforementioned setups for analogue simulations of quantum gravity\cite{pesao1,pesao2,pesao3,pesao4}.

In this work, we focus instead in the simulation of the spacetime metric of a bouncing black hole and to this end we use a different superconducting-circuit setup, consisting of a dc-SQUID array embedded in a transmission line. In this eventual experimental setup, the speed of propagation of a quantum field could be modulated through an external magnetic field threading the SQUIDs, a fact that could be used to mimic the speed of propagation in a curved spacetime \cite{squid1,squid2,squid3} including black holes \cite{squid6,squid4}. We apply these techniques to the spacetime metric considered in \cite{Barceló_2015,mutinity}, where the standard Schwarzschild metric is modified to accommodate a bounce, namely a black hole to white hole transition. In this way, the same theoretical framework and experimental setup that is used to simulate black holes, which in principle gives rise to a horizon and a singularity, could be used as well to simulate a scenario in which the process stops and reverts at some point, enabling the possibility of analyzing for instance the robustness of Hawking radiation \cite{garcia2023hawking, barceloanalogue}.

\section{Theoretical framework}

The standard Schwarzschild metric -natural units $G=c=1$ and Schwarzschild coordinates- of a stellar body is given by
\begin{equation}
ds^2=-\left(1-\frac{2M}{r}\right)dt^2+\frac{1}{1-\frac{2M}{r}}dr^2+r^2d\theta^2+\,r^2\sin^2\theta d\phi^2,
\end{equation}
in the outside, while inside the collapsing star:
\begin{eqnarray}
ds^2&=&-\frac{1}{4}\left(3\sqrt{1-\frac{2M}{r_\star}}-\sqrt{1-\frac{2Mr^2}{r_\star^3}}\right)^2dt^2+\nonumber\\ & &\frac{1}{1-\frac{2Mr^2}{r_\star^3}}dr^2+r^2d\theta^2+r^2\sin^2\theta d\phi^2,
\end{eqnarray}
where $M$ is the mass of the star and $r_\star=r_\star(t)$ its radius.

We can switch to coordinates that result more convenient, those of Gullstrand-Painlevé\cite{painleve, gullstrand}. This coordinate system is defined as the one associated to a freely falling observer from infinity, characterized by flat spatial sections and a time coordinate equal to the proper time of this observer.

In this new coordinate system, the metric takes the following form:
\begin{equation}
ds^2=-\left(c^2-v^2\right)dt^2-2vdrdt+dr^2+r^2d\Omega^2
\end{equation}
with $c=1$ and $v=v(t, r)$ a piecewise function of the coordinates, defined differently in each collapse period.

In this work we consider a modification of the metric above to accommodate a period of collapse and a period of expansion of the star, centered at $t=0$, separated by a bounce period that smoothly joins the two metrics with a duration $t_b$, i.e., with $t\in[-\sfrac{t_b}{2}, \sfrac{t_b}{2}]$ \cite{collapse, mutinity}.

For $t<-\sfrac{t_B}{2}$, the value of $v(t, r)$ is:
\begin{equation}
v=\left\{\begin{matrix}
-\sqrt{\frac{2M}{r}} & \mathrm{if} & r>r_\star(t) \\
-\sqrt{\frac{2Mr^2}{r_\star^3(t)}} & \mathrm{if} & 0<r<r_\star(t)
\end{matrix}\right.\label{in}
\end{equation}

Whereas for $t>\sfrac{t_B}{2}$ it is given by the opposite:
\begin{equation}
v=\left\{\begin{matrix}
\sqrt{\frac{2M}{r}} & \mathrm{if} & r>r_\star(t) \\
\sqrt{\frac{2Mr^2}{r_\star^3(t)}} & \mathrm{if} & 0<r<r_\star(t)
\end{matrix}\right.\label{exit}
\end{equation}

For the time $-\sfrac{t_B}{2}<t<\sfrac{t_B}{2}$, quantum gravity phenomena emerge, for which we do not yet have a theory, so we cannot predict the form of the metric in this interval. During this short period of time, a time-symmetric bounce occurs, interpolating between the collapse and expansion regions.

Since this is the collapse of a star without pressure in its interior, the radius of the star varies as if it were a particle in free fall. Since we are in Gullstrand-Painlevé coordinates, the rate at which the star's surface falls is simply $v$ evaluated at the boundary:
\begin{equation}
\frac{dr_\star}{dt}=-\sqrt{\frac{2M}{r_\star}}\to\frac{2}{3}r_\star^{\sfrac{3}{2}}=-\sqrt{2M}t\to r_\star^3=\frac{9M}{2}t^2
\end{equation}
where integration constants have been adjusted so that $r_\star=0$ at time $t=0$. The collapsing star becomes a black hole when $r_\star=2M$, i.e., at $t=\pm\frac{4M}{3c}$.

The effective speed of light in the radial direction as a function of the coordinates can be obtained by solving the quadratic equation $ds^2=0$, leading to:
\begin{equation}
\tilde c=\frac{dr}{dt}=\frac{2v\pm\sqrt{4v^2+4\left(c^2-v^2\right)}}{2}=v\pm c,
\end{equation}
where different possibilities for the signs give rise to either motion in the direction of $v$ or opposing it. In the pedagogically useful language of the river model of black holes \cite{hamilton2008river}, this would correspond to light moving ``downstream'' or ``upstream'', as we will see in more detail in the Section Simulation Proposal.

\section{Quantum simulation of curved spacetime in SQUID arrays}

A \textit{superconducting quantum interference device} (SQUID) consists of two Josephson junctions \cite{josephson} connected in parallel in a superconducting circuit, with a gap through which a magnetic field can be applied \cite{you_2011}.

A dc-SQUID array is a one-dimensional metamaterial formed by multiple dc-SQUIDs connected in series in a straight line. This array can be used as a waveguide for an electromagnetic field, whose propagation speed depends on the magnetic flux applied to each SQUID as follows \cite{simoen}:
\begin{equation}
c(x)=\frac{1}{\sqrt{CL(\phi(x))}}
\end{equation}
where $c(x)$ is the effective speed of light induced at each point of the SQUID array and $C$ and $L$ are the capacitance and inductance of the system per unit length, with:
\begin{equation}
L(\phi(x))=\frac{\phi_0}{4\pi I_c\left|\cos\left(\pi\frac{\phi(x)}{\phi_0}\right)\right|\cos\psi}
\end{equation}
where $\phi(x)$ is the magnetic flux at each point of the SQUID array, $\phi_0$ is the magnetic flux quantum, given by $\phi_0=\sfrac{h}{2e}$, $I_c$ is the critical current of the Josephson junction, and $\psi$ is the phase difference between each SQUID. In the linear regime, we assume $\cos\psi\approx1$. More simply, the speed of light is:
\begin{equation}
c^2(x)=c_0^2\left|\cos\left(\pi\frac{\phi(x)}{\phi_0}\right)\right|
\end{equation}
where:
\begin{equation}
c_0=c(\phi=0)=\ell\sqrt{\frac{4\pi I_c}{\phi_0 C}}
\end{equation}
with $\ell$ being the length of the SQUID array.

The experimental setup on which we base our model consists of a one-dimensional array of SQUIDs in which we can induce an effective speed of light at each point, which can be varied in time during the experiment, therefore inducing an effective speed of light at each point of a 1+1 dimensional spacetime. This is analogous to what happens in General Relativity, where the effective speed of light is different at each point in spacetime due to the curvature encoded in the metric. Therefore, we can understand this setup as an effective 1+1 D curved spacetime with an effective speed of light $\tilde{c}(x)$, as long as the magnetic flux obeys \cite{squid3}:
\begin{equation}
\frac{\pi\phi^{AC}}{\phi_0}=\arccos\left(\cos\left(\frac{\pi\phi^{DC}}{\phi_0}\right)\Tilde{c}^2\right)-\frac{\pi\phi^{DC}}{\phi_0}\label{flujovel}
\end{equation}
where:
\begin{equation}
\tilde{c}^2=\left|\sec\left(\frac{\pi\phi^{DC}}{\phi_0}\right)\right|\left|\cos\left(\frac{\pi\phi}{\phi_0}\right)\right|,
\end{equation}
and the flux has been split into DC and AC parts $\phi=\phi^{AC}+\phi^{DC}$.

\section{Simulation Proposal}\label{sec:simula}

Let us consider both the negative and positive signs of $c=\pm1$. The negative sign corresponds to light falling radially, while the positive sign corresponds to light moving away from the star. When $v$ and $c$ have opposite signs -black hole and moving away / white hole and infalling-, the light travels \textit{upstream}, while if $v$ and $c$ have the same sign -black hole and falling / white hole and outward motion-, the light travels \textit{downstream}.

When substituting into equation \ref{flujovel}, we must consider that the arccosine function has a domain in the interval $[-1, 1]$, so it is necessary to introduce a constant magnetic flux $\phi^{DC}$ to avoid superluminal speeds in the laboratory -which of course cannot be generated- when we want to simulate that $v$ and $c$ go in the same direction, or when we are too close to the black hole singularity. The effect of $\phi^{DC}$ is to effectively reduce the speed of light of vacuum in the simulated spacetime \cite{squid3}.

For the metric before the bounce and with the light moving away from the star, the AC magnetic flux needed to simulate the speed of light $\tilde{c}=v+c$ is:
\begin{equation}
    \phi^{AC}=\frac{\phi_0}{\pi}\arccos\left(\cos\left(\frac{\pi}{\phi_0}\phi^{DC}\right)\left(v+c\right)^2\right)-\phi^{DC}
\end{equation}

Then, for $r>r_\star$, the total flux $\phi$ is:
\begin{equation}
\phi=\frac{\phi_0}{\pi}\arccos\left(\cos\left(\frac{\pi}{\phi_0}\phi^{DC}\right)\left(-\sqrt{\frac{2M}{r}}+1\right)^2\right)\label{phiBHout}
\end{equation}
and, for $r<r_\star$:
\begin{eqnarray}
\phi&=&\frac{\phi_0}{\pi}\arccos\left(\cos\left(\frac{\pi}{\phi_0}\phi^{DC}\right)\left(-\sqrt{\frac{2Mr^2}{r_\star^3(t)}}+1\right)^2\right)\nonumber\\
&=&\frac{\phi_0}{\pi}\arccos\left(\cos\left(\frac{\pi}{\phi_0}\phi^{DC}\right)\left(-\sqrt{\frac{4r^2}{9t^2}}+1\right)^2\right)\label{phistarout}
\end{eqnarray}

As we have explained, since the arccosine function has a domain between -1 and 1, it is necessary to adjust $\phi^{DC}$ so that the argument does not fall outside its domain. However, we cannot do this for all points in the $(t, r)$ plane, as the argument of function \ref{phiBHout} diverges when $(t, r)\to(0, 0)$. Nevertheless, we can exclude from this study the points where quantum gravity processes emerge, between $-\sfrac{t_B}{2}$ and $\sfrac{t_B}{2}$, as we do not really know the metric in this region, and calculate the value of $\phi^{DC}$ so that the arccosine is well defined outside this region. The value of $\phi^{DC}$ that meets this requirement can be calculated so that, exactly at the boundary, with $t=\pm\sfrac{t_B}{2}$, and $r=r_\star\left(\pm\sfrac{t_B}{2}\right)=\frac{1}{2}\sqrt[3]{9Mt_B^2}$, the argument inside the arccosine equals exactly 1:
\begin{eqnarray}
    & &\cos\left(\frac{\pi}{\phi_0}\phi^{DC}\right)\left(1-\sqrt{\frac{4M}{\sqrt[3]{9Mt_B^2}}}\right)^2=1  \longrightarrow \nonumber\\ &\phi^{DC}&=\frac{\phi_0}{\pi}\arccos\left(\frac{1}{\left(1-2\sqrt[3]{\frac{M}{3t_B}}\right)^2}\right)
\end{eqnarray}

\begin{figure}
\includegraphics[width=0.45\textwidth]{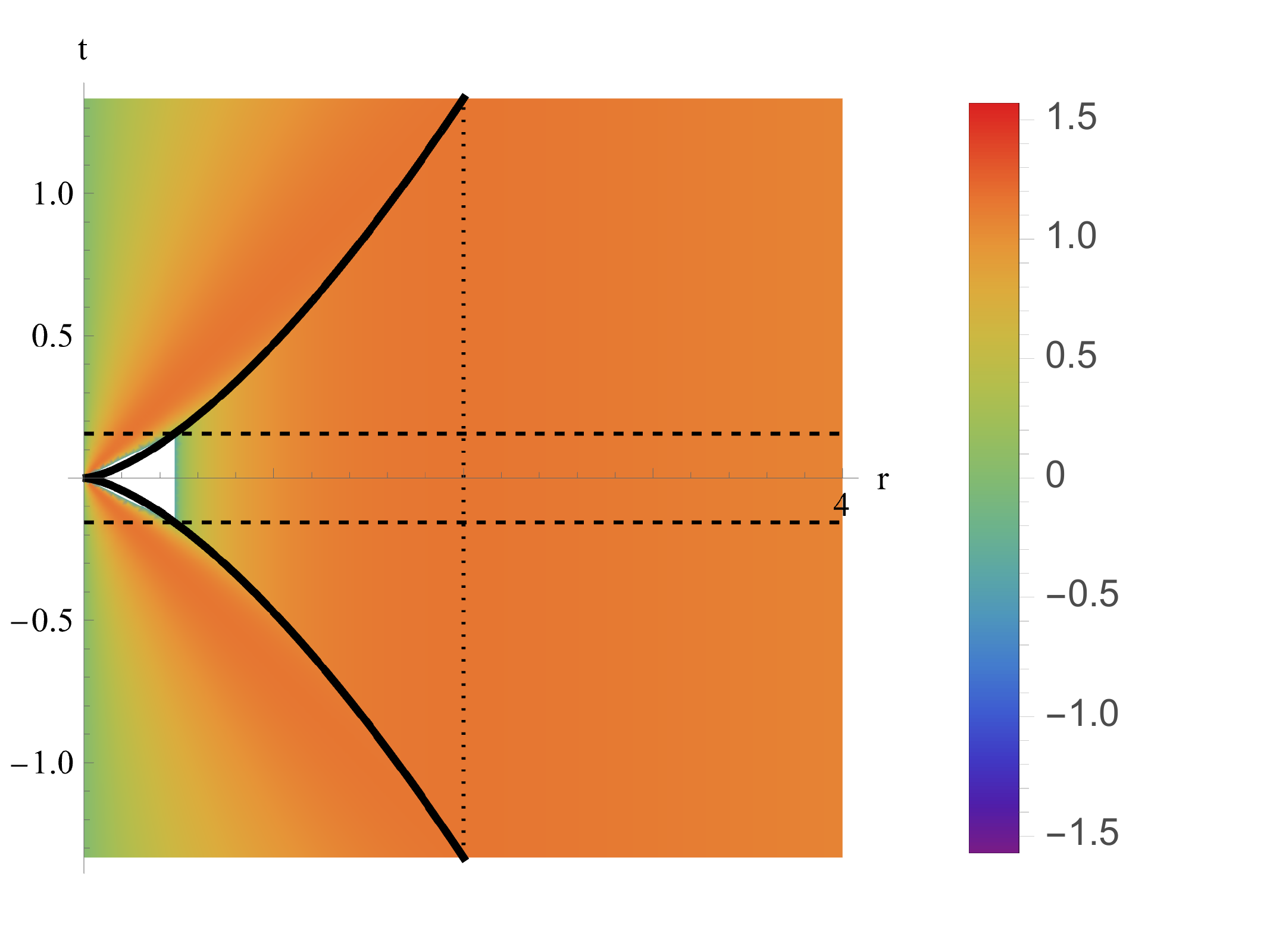}
\caption{Applied magnetic flux, in units of $\sfrac{\hbar}{e}$, needed to induce a propagation speed that simulates the speed of light upstream in a Schwarzschild spacetime, with the metric of equations \ref{in} and \ref{exit}, as a function of $r$ and $t$, in units of $M$. The solid black line represents $r_\star(t)$. The horizontal dashed lines represent $\pm\sfrac{t_B}{2}$. The vertical dashed line represents the black hole event horizon.}
\label{fig:1}
\end{figure}

Substituting these values in \ref{phiBHout} for $r>r_\star$ gives:
\begin{eqnarray}
\frac{\pi}{\phi_0}\phi^{AC}&=&\arccos\left(\left(\frac{1-\sqrt{\frac{2M}{r}}}{1-2\sqrt[3]{\frac{M}{3t_B}}}\right)^2\right)\nonumber\\ &-&\arccos\left(\frac{1}{\left(1-2\sqrt[3]{\frac{M}{3t_B}}\right)^2}\right)\label{phiBHoutprime}
\end{eqnarray}
and, in \ref{phistarout}, for $r<r_\star$:
\begin{equation}
\begin{split}
    \frac{\pi}{\phi_0}\phi^{AC}=&\arccos\left(\left(\frac{1-\sqrt{\frac{4r^2}{9t^2}}}{1-2\sqrt[3]{\frac{M}{3t_B}}}\right)^2\right)\\-& \arccos\left(\frac{1}{\left(1-2\sqrt[3]{\frac{M}{3t_B}}\right)^2}\right)\label{phistaroutprime}
\end{split}
\end{equation}

Again, since arccosine is defined only up to 1, in the last term it must be $3t_B<M$ or, reconstructing the units, $3t_BG<Mc^3$. In principle, this is not a fundamental limit but a criterion for being able to compute the magnetic flux at any point in the simulated spacetime, except within a small region during the bounce, which was already excluded from the simulation. However, for the mass values expected for these processes, i.e. for black holes with Planck mass, this computational limit translates to the physically meaningful conclusion of the bounce time having to be at most one-third of the Planck time. 

Now, for the metric after the bounce, and with light approaching the star, the magnetic flux needed to simulate the speed of light $\tilde{c}=v-c$, with $\phi^{DC}=0$, is the same as in the previous cases, only with $v$ and $c$ having opposite signs. However, since the function is squared, this sign change does not alter it, so the magnetic flux needed to simulate the system is exactly the same as in the previous case, described by equations \ref{phiBHoutprime} and \ref{phistaroutprime}, and represented in Fig. \ref{fig:1}.

For the metric before the bounce and with light approaching the star, the alternating current magnetic flux needed to simulate the speed of light $\tilde{c}=v-c$ is:
\begin{equation}
\phi^{AC}=\frac{\phi_0}{\pi}\arccos\left(\cos\left(\frac{\pi}{\phi_0}\phi^{DC}\right)\left(v-c\right)^2\right)-\phi^{DC}
\end{equation}

Then, considering the total flux $\phi=\phi^{AC}+\phi^{DC}$, we have for $r>r_\star$:
\begin{equation}
\phi=\frac{\phi_0}{\pi}\arccos\left(\cos\left(\frac{\pi}{\phi_0}\phi^{DC}\right)\left(-\sqrt{\frac{2M}{r}}-1\right)^2\right) \label{phiBHin}
\end{equation}
\begin{figure}
\includegraphics[width=0.45\textwidth]{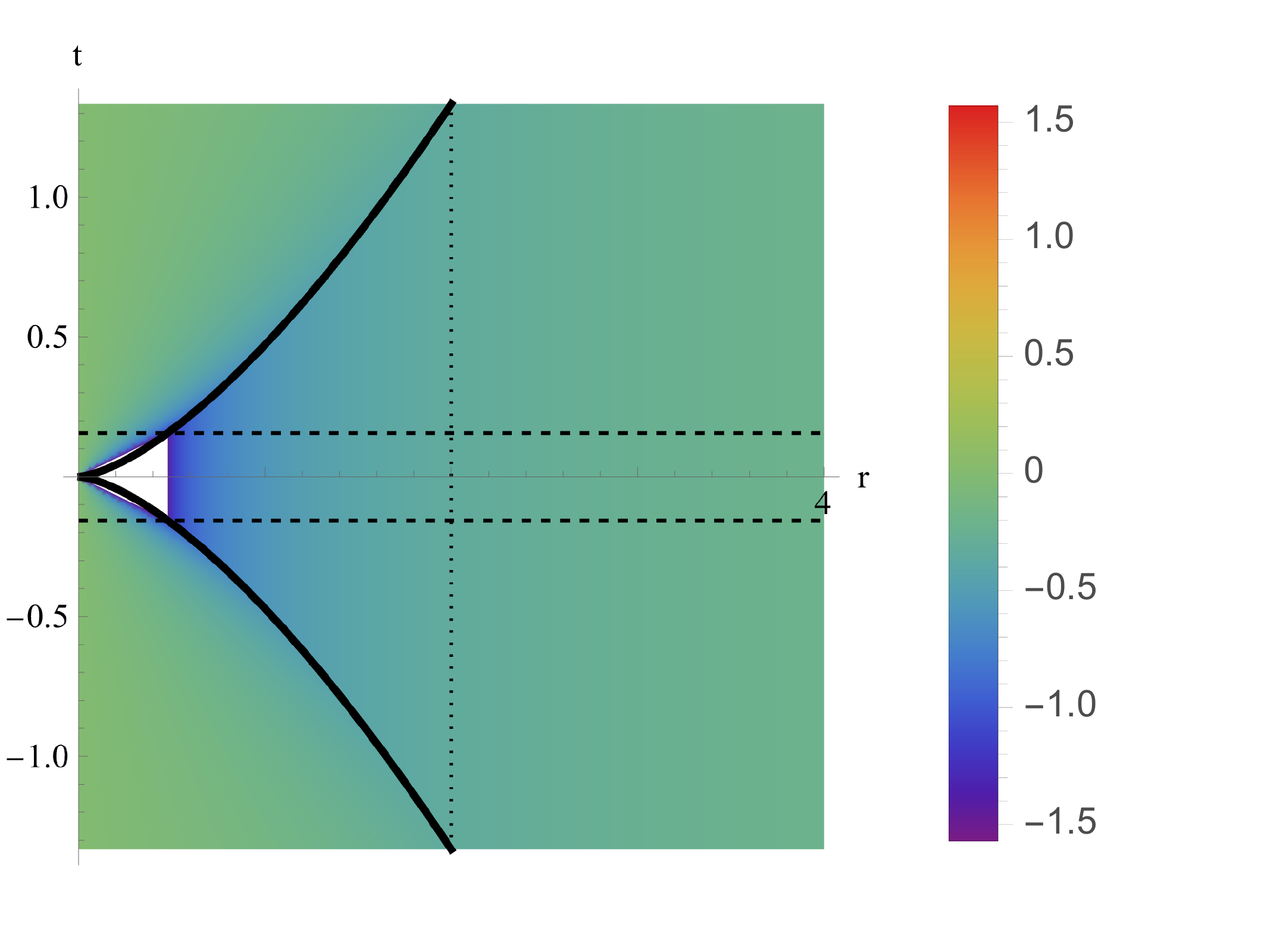}
\caption{Applied magnetic flux, in units of $\sfrac{\hbar}{e}$, needed to induce a propagation speed that simulates the speed of light downstream in a Schwarzschild spacetime, with the metric of equations \ref{in} and \ref{exit}, as a function of $r$ and $t$, in units of $M$. The solid black line represents $r_\star(t)$. The horizontal dashed lines represent $\pm\sfrac{t_B}{2}$. The vertical dashed line represents the black hole event horizon.}
\label{fig:2}
\end{figure}
and, for $r<r_\star$:
\begin{eqnarray}
\phi&=&\frac{\phi_0}{\pi}\arccos\left(\cos(\frac{\pi\phi^{DC}}{\phi_0})\left(-\sqrt{\frac{2Mr^2}{r_\star^3(t)}}-1\right)^2\right)\nonumber\\
&=&\frac{\phi_0}{\pi}\arccos\left(\cos(\frac{\pi\phi^{DC}}{\phi_0})\left(-\sqrt{\frac{4r^2}{9t^2}}-1\right)^2\right)\label{phistarin}
\end{eqnarray}

Similarly to what we did in the previous case, we calculate the direct current magnetic field needed to simulate the entire region outside the zone $-\sfrac{t_B}{2}<t<\sfrac{t_B}{2}$ as:
\begin{eqnarray}
& &\cos\left(\frac{\pi}{\phi_0}\phi^{DC}\right)\left(-\sqrt{\frac{4M}{\sqrt[3]{9Mt_B^2}}}-1\right)^2=1\longrightarrow \nonumber\\ & &\phi^{DC}=\frac{\phi_0}{\pi}\arccos\left(\frac{1}{\left(1+2\sqrt[3]{\frac{M}{3t_B}}\right)^2}\right)
\end{eqnarray}

Substituting these values in \ref{phiBHin} for $r>r_\star$:
\begin{eqnarray}
\frac{\pi}{\phi_0}\phi^{AC}&=&\arccos\left(\left(\frac{1+\sqrt{\frac{2M}{r}}}{1+2\sqrt[3]{\frac{M}{3t_B}}}\right)^2\right)\nonumber\\ &-&\arccos\left(\frac{1}{\left(1+2\sqrt[3]{\frac{M}{3t_B}}\right)^2}\right)\label{phiBHinprime}
\end{eqnarray}
and, in \ref{phistarin}, for $r<r_\star$:
\begin{equation}
   \frac{\pi}{\phi_0}\phi^{AC}=\arccos\left(\left(\frac{1+\sqrt{\frac{4r^2}{9t^2}}}{1+2\sqrt[3]{\frac{M}{3t_B}}}\right)^2\right)- \arccos\left(\frac{1}{\left(1+2\sqrt[3]{\frac{M}{3t_B}}\right)^2}\right)\label{phistarinprime}
\end{equation}

For the same reason as before, for the metric after the bounce and with the light moving away from the star, the alternating current magnetic flux required to simulate the speed of light $\tilde c=v+c$ is the same as in the previous cases, described by equations \ref{phiBHinprime} and \ref{phistarinprime}, and represented in Fig.\ref{fig:2}.

In general, the effect of the simulated curved geometry translates into a modification of the equations of motion of the propagating field, which thus acquires a particular phase shift. These phase shifts can be measured with state-of-the-art superconducting circuit technology\cite{squid1, PhysRevA.90.052113}.

\section{Discussion}

For \textit{upstream} light, we find maximum applied magnetic fluxes to simulate the system at $r=2M$ outside the collapsing body and at $r=\frac{3}{2}t$ inside it, as long as the black hole has already formed, i.e., with $r_\star$ smaller than the Schwarzschild radius. If the black hole had not yet formed or had disintegrated after the bounce, we would only find a single maximum of the applied magnetic flux on the surface of the star, at $r=r_\star(t)$. At the points $(r, t)=\left(2M, \pm\frac{4M}{3}\right)$ where the star becomes a black hole, these three maxima coincide.

For \textit{downstream} light, we only find the minima of the magnetic flux -maxima of the absolute value- on the surface of the collapsing body $r=r_\star(t)$, regardless of whether it has become a black hole or not. 

At these points, we have $\phi=\pm \phi_0/2$ -which means $\tilde{c}=0$ and infinite inductance. Therefore, with these critical values, quantum fluctuations would appear because of the very high impedance of the electromagnetic environment. If a large region of the array is close to this limit, this could lead to large fluctuations in the superconducting phase $\psi$, breaking our approximation $\cos \psi \simeq 1$ and preventing the system from being in the superconducting phase \cite{squid_wormhole_PhysRevD.94.081501, exotic_spacetimes_Sab_n_2018,squid_array_Haviland2000}. Therefore, we should try to avoid it by keeping as few SQUIDs as possible close to this limit, ideally a single SQUID.
Note that in a potential experimental implementation, these would be points of great interest, as photon pairs would be produced on them in the \textit{SQUID array}\cite{squid5, casimir}, analogous to the production of Hawking radiation in high-gravity environments\cite{squid6}. In Fig.\ref{fig:1} we see that the flux should be close to the critical value in almost all spacetime outside the horizon, while in Fig. \ref{fig:2} we see that the flux gets close to the critical value only inside the horizon in a small spacetime region around the boundary. This suggests that the latter scheme -downstream light- might be more feasible to realize in an experiment: the quantum fluctuations of the phase could be contained in a small region of the array while the system remains in the superconducting phase regime. 

\section{Summary and conclusions}
We propose an in-principle analog quantum simulation of a spacetime inspired in loop quantum gravity models \cite{CRStephens_1994,Barceló_2015,mutinity,PhysRevD.106.024014}: the collapse and bounce of a star from infinity, or in other words, a black hole which bounces and transforms into a white hole after reaching a critical Planck-scale size, therefore avoiding the singularity and the information loss problem. In this spacetime, we consider a massless scalar field, whose propagation speed is modified by the curvature. We can simulate a radial section of this system by using a one-dimensional SQUID array, in which we can modify the speed of propagation of a quantum electromagnetic field by means of an external magnetic field threading the SQUIDs. We consider both infalling and outfalling radiation, giving rise to two different scenarios: downstream and upstream radiation. We compute the magnetic flux profile required by the simulation in both cases and find that the former is more experimentally suitable, since critical values of the magnetic flux appear only in small regions of the simulated spacetime, suggesting that the corresponding quantum fluctuations of the superconducting phase can be contained in a small region of the array without leaving the superconducting regime in the whole array. Indeed, these quantum fluctuations can be interpreted as quantum corrections to the classical spacetime and therefore as an analogue to quantum backreaction. Moreover, a future experimental realization of this system, when the technology is capable to realize this proposal, could allow us to explore the dynamics of quantum fields inside black holes, a problem that remains open today and whose study could lead to a deeper understanding of fundamental problems such as the information loss.

\section*{Acknowledgements}
CS acknowledges financial support through the Ramón y Cajal Programme (RYC2019-028014-I).

\section*{Author contributions statement}

CS designed the project. MDM developed and executed the project, performed the computations and figures, and wrote the first version of the manuscript. Both authors reviewed the manuscript.

\section*{Data availability}

The authors declare that the data supporting the findings of this study are available within the paper.

\section*{Additional information}

\textbf{Competing interests} 

The authors declare no competing interests.  





\end{document}